\documentclass[aps,pra,twocolumn,showpacs]{revtex4}
\usepackage{graphicx}\usepackage{amssymb}\usepackage{hyperref}
\begin{document}
\title{Superfluidity of two-dimensional excitons in flat and harmonic traps}
\author{Yu.E.~Lozovik$^a$, I.L.~Kurbakov$^a$, and Magnus~Willander$^b$}
\affiliation{$^a$ Institute of Spectroscopy, Russian Academy of Sciences,
142190 Troitsk, Moscow region, Russia\\
$^b$ Institute of Science and Technology (ITN), Link\"oping University,
SE-581 83, Link\"oping, T\"appan 6177, Sweden}
\date{\today}
\begin{abstract}
Superfluid exciton density and superfluid transition (cross\-over) temperature
are calculated for 2D excitons in large-size flat and harmonic traps. A
generalized local density approximation for the Kosterlitz-Thouless theory is
developed.
\end{abstract}
\maketitle
\section{Introduction}
Two-dimensional (2D) indirect exciton systems in coupled quantum wells (CQWs)
\cite{LY} (or in single QW in strong normal electric field \cite{SQW}), in
which electrons and holes are spatially separated, at low temperatures can
possess condensation and superfluidity \cite{LY,SQW,SFB,LB,BEC&SF,LV} (or
crystallization in some region of phase diagram \cite{LV,cr,MCcr}), persistent
electric currents and quasi-Josephson effects \cite{YKSP}, as well as
interesting coherent optical phenomena \cite{opt}. There is a tremendous
progress now in the observation of collective exciton properties in CQWs (see
reviews \cite{revM,revT,revB,revS}).

Due to the divergence of phase fluctuations, Bose condensation of excitons
(possible in 3D systems \cite{KK}) in {\it extended} 2D exciton systems at
$T\ne0$ is impossible \cite{r1580383}. But at low temperatures the system can
possess quasi-long (power-law) off-diagonal order \cite{r1550080} and
superfluidity, which are destroyed by vortex pair dissociation and free
vortex formation at Kosterlitz-Thouless (KT) temperature \cite{Berezinskii,%
jc061181,jc071046}. But Bose condensation is possible in {\it finite} 2D
systems \cite{2DBEC}. Besides, the trapping can increase the exciton density
at fixed laser pumping, which leads to the growth of Bose condensation
temperature. Several types of traps were proposed and experimentally realized:
(i) "natural" traps created by (one-monolayer) well width fluctuations
\cite{Toldnew,na417047,b6400413}; and regular, controlled traps such as (ii)
laser-induced traps formed inside the annulus of laser pump \cite{rl962202},
(iii) traps formed by an inhomogeneous deformation created by a needle
stressing on the sample \cite{rl970103,ss134037}, (iv) mesa created by
engineering width of layers or the barrier in CQWs, and (v) different kinds
of electrostatic traps \cite{revT,a0990604,mj360940,b7400409}. The interesting
optical properties of exciton collective state are observed in luminescence
measurements \cite{Toldnew,na417047,b6400413,rl962202,b7400409,ringS,liqu,B}
(see also \cite{revM,revT,revB,revS}).

In connection with the discussion above essential problems arose concerning
the nature of exciton collective state in 2D traps: (i) How the phase
fluctuations and vortices manifest itself in sufficiently large but finite
traps? (ii) How the type of the confining potential influences on the
superfluid density spatial distribution and its evolution with the
temperature? These problems the present paper are mainly addressed to.

We describe quantitatively the superfluidity of 2D CQW excitons in 2D
large-size flat trap by means of KT theory. For the investigation of the
exciton superfluidity in 2D large-size harmonic trap we employ the local
density approximation \cite{LDAcorr} and develop more correct generalized
local density approximation for KT theory.

\section{Superfluidity of 2D excitons in a large-size harmonic trap: The
local density approximation}\label{LDA}
As the first step for study of 2D CQW exciton superfluidity in a large-size
harmonic trap we use the local density approximation (LDA). LDA is valid if
confining potential of the trap varies sufficiently {\it slowly} on {\it all}
microscopical characteristic scales of the problem, particularly, related to
phonon and vortex excitations in the exciton system. Therefore, in LDA exciton
superfluid properties near the trap point ${\bf R}$ coincides with properties
of the homogeneous {\it infinite} superfluid with total density $n_t^{\infty}$
being equal to the (total) exciton density $n_t({\bf R})$ in the trap in the
point ${\bf R}$.

A homogeneous infinite exciton system at fixed temperature $T$ is a superfluid
at high total densities $n_t^{\infty}$ but it is a normal fluid at low
$n_t^{\infty}$. As total exciton density $n_t^{\infty}$, when decreasing, goes
through a critical one $\tilde n_t^{\infty}$ corresponding to exciton
temperature $T$ the superfluid KT transition is occured in the system and
superfluid exciton density $n_s^{\infty}$ vanishes by the {\it universal jump}
$\tilde n_s^{\infty}=2mT/\pi\hbar^2$ \cite{rl391201} ($m$ is the exciton
mass).

So, the trapped excitons in LDA are superfluid (normal) ones at the trap point
${\bf R}$ at which total exciton density $n_t({\bf R})$ is greater (less) than
the critical one $\tilde n_t^{\infty}$, corresponding to exciton temperature
$T$. Thus, in LDA at the sufficiently low temperatures at which the critical
total exciton density $\tilde n_t^{\infty}$ is smaller than maximal one
$n_t^m=n_t(0)$ the exciton system in a {\it symmetrical} harmonic trap is
divided into a {\it superfluid circle} $R<R_s$ of a radius $R_s$ and {\it
normal ring} $R_s<R<\bar L$ surrounding the circle ($\bar L$ is the
Thomas-Fermi (TF) radius). In the superfluid circle the total exciton density
$n_t(R)$ exceeds a critical value $\tilde n_t^{\infty}$ corresponding to
exciton temperature $T$. But the superfluid density $n_s(R)$ is proved (see
details in \cite{jc061181,rl391201}) to be higher than the universal jump
$\tilde n_s^{\infty}$ for this temperature. In opposite, in the normal ring
total exciton density $n_t(R)$ is smaller than $\tilde n_t^{\infty}$ and
superfluid density is equal to zero $n_s(R)\equiv0$. In the circumference
$R=R_s$ which is the boundary line of the superfluid circle and normal ring
the {\it local} KT transition takes place on which the onset of pair vortex
dissotiation and free vortex formation occurs. At the circumference $R=R_s$
the superfluid density jumps from $\tilde n_s^{\infty}$ to zero.

As the temperature grows critical total exciton density $\tilde n_t^{\infty}$
increases. So, the superfluid circle shrinkes and the normal ring thickens. At
the temperature $\bar T_c^{\infty}$, at which the critical total density
becomes equal to {\it maximal} total density in the whole trap $n_t^m=n_t(0)$,
the superfluid circle disappears and the trapped exciton system undergoes a
{\it global} transition to the normal state.

\section{Superfluidity of 2D excitons in a large-size flat trap}\label{FT}
In 2D homogeneous (flat) trap of sufficiently large size $L$ the true,
vortex-renormalized (VR), superfluid exciton density $n_s$ takes into account
the renormalization ("screening") by all vortex pairs in the system, i.e., by
vortex pairs with the separation less than $L$. So, VR superfluid density
$n_s$ is expressed through "local", vortex-unrenormalized (VU), superfluid
density $n_l$ as follows
\begin{equation}\label{ns}
n_s=n_l/\epsilon(L),
\end{equation}
where $\epsilon(r)$ is vortex pair dielectric function \cite{jc061181,%
b3203088,VR}. In superfluid phase where there is a rare vortex pair gas the
quantity $\epsilon(L)$ according to KT model \cite{jc061181,b3203088} obeys
the equation
\begin{equation}\label{dex/dx}
\frac{d\epsilon(x)}{dx}=C(a)e^{4x-a\int_0^xdx'/\epsilon(x')}
\end{equation}
with the boundary condition
\begin{equation}\label{e0=1}
\epsilon(0)=1.
\end{equation}
Here $x=\ln(L/r_0)$, $r_0$ is vortex core diameter (in CQWs $r_0$ is of order
of the distance between excitons \cite{MC}),
\begin{equation}\label{a}
a=2\pi\hbar^2n_l/mT,
\end{equation}
and $C(a)$ is a (nonuniversal) constant depending on microscopical exciton
superfluid properties; in 2D $X$-$Y$ model it is equal to
$C(a)=\pi^2ae^{-\pi a/2}$.

According to our numerical analysis of Eqs. (\ref{ns})-(\ref{e0=1}), at low
temperatures $T$ (i.e., at large $a$; see Eq. (\ref{a})) VR superfluid density
$n_s$ is close to VU one $n_l$. As $T$ grows the ratio $n_s/n_l$ decreases. At
a critical temperature $T_c^L$ in sufficiently large traps ($\ln(L/r_0)\gg1$)
there is the {\it crossover} where rather sharp disappearance of the VR
superfluid density takes place. The disappearance corresponds to the onset of
pair vortex dissociation and {\it free vortex} formation \cite{jc061181},
superfluid state being destroyed.

\begin{figure}[t]
\includegraphics[width=8.65cm,height=5.1cm]{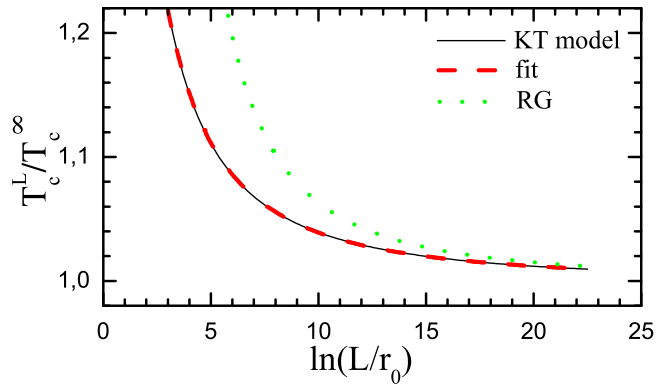}
\vskip -4mm
\caption{\small The dependence of the superfluid crossover temperature $T_c^L$
of 2D CQW excitons in the flat trap on its size $L$
($T_c^{\infty}\equiv\left.T_c^L\right|_{L=\infty}$). Thin solid line is the
numerical intergation of the equation for vortex pair dielectric function
$\epsilon(L)$ in KT model, dash line is the fit
$\tilde\epsilon^{\infty}/\tilde\epsilon^L$ (according to Eq. (\ref{eLcr})),
and dot line is the Kosterlitz's renormalization group calculation.}
\end{figure}

Numerical integrating Eqs. (\ref{dex/dx}), (\ref{e0=1}) yield the following
expression for the superfluid crossover temperature for CQW excitons in 2D
flat trap of size $L$:
\begin{equation}\label{_TcL}
T_c^L=\frac{\pi\hbar^2n_l(T_c^L)}{2m\tilde\epsilon^L},
\end{equation}
where $n_l(T_c^L)$ is the VU superfluid density of 2D excitons at the
crossover (at $T=T_c^L$) and the quantity $\tilde\epsilon^L$ is calculated
numerically. The fit
\begin{equation}\label{eLcr}
\tilde\epsilon^L=\left(1-\frac{\pi^2b^2}{(\ln(L/r_0)+\Delta)^2}\right)
\tilde\epsilon^{\infty}
\end{equation}
coincides excellently with the numerical calculation at large $L$ (see below
and Fig. 1); in Eq. (\ref{eLcr}) $\tilde\epsilon^{\infty}$ is the critical
vortex pair dielectric constant in 2D infinite superfluid at KT transition,
$b$ and $\Delta$ are nonuniversal quantities which, as
$\tilde\epsilon^{\infty}$, depend on microscopical exciton system properties.
In 2D $X$-$Y$ model we obtain $\tilde\epsilon^{\infty}=1.134(9)$,
$\Delta=2.(93)+\tilde\Delta$ and $b=0.8(00)$. The quantity
$\tilde\Delta=\tilde\Delta(L)$ determines the correctness of KT model, so,
within KT model one must assume that $\tilde\Delta=0$. The numerical
calculation of the width of superfluid crossover region, in which vortex pairs
should begin to dissociate, yields the following estimation
$e^{\tilde\Delta}\sim1$.

From Eqs. (\ref{_TcL}), (\ref{eLcr}) we conclude that the approach used takes
into account the {\it second} order on the logarithmically small quantity
$1/(\ln(L/r_0)+\Delta)$. Thus, our calculation is correct in the {\it
logarithmic} approximation ($\ln(L/r_0)\gg1$), i.e., for sufficiently large
traps.

For the halfwidth of superfluid crossover region we obtain (see Eqs.
(\ref{_TcL}) and (\ref{eLcr}))
\begin{equation}\label{DTcL}
\Delta T_c^L\sim T_c^L\frac{\pi^2b^2}{(\ln(L/r_0)+\Delta)^3}.
\end{equation}
Thus, the model adopted does not take into account effects in the crossover
region (because $\Delta T_c^L$ depends on the third order of
$\ln(L/r_0)\gg1$). In the crossover temperature region $\Delta T_c^L$ the
number of free vortices inside the trap area $L^2$ increases by $e$ times,
whereas at $T=T_c^L$ there is (in average) of order of one free vortex.

Note that the result (\ref{DTcL}) for 2D differs essentially from the case of
3D Bose condensed systems where $\Delta T_c^L$ depends on a power of $L$.

\begin{figure}[t]
\includegraphics[width=8.65cm,height=5.1cm]{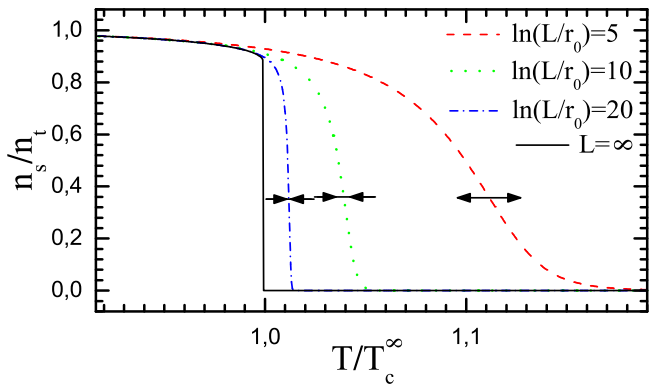}
\vskip -4mm
\caption{\small The dependence of the superfluid density of 2D CQW excitons in
the flat trap of different size $L$ on the ratio of exciton temperature $T$ to
the superfluid transition temperature $T_c^{\infty}$ of 2D infinite systems.
The regions of the superfluid crossovers on which the superfluid density
universal jump is smeared out are shown by arrows.}
\end{figure}

For the completeness of the calculation of VR superfluid density $n_s$ and
superfluid crossover temperature $T_c^L$ in the flat trap one must calculate
the dependence of VU superfluid density $n_l$ on temperature $T$ and total
density $n_t$ (see Eqs. (\ref{ns}), (\ref{a}) and (\ref{_TcL})). This
calculation is realized in Refs. \cite{MC,TT}. In this Letter we make a
simplification. Namely, for spatially indirect CQW excitons the strong
dipole-dipole repelling results in a small {\it thermal} depletion of their
{\it VU superfluid density} (not the condensate) throughout the superfluid
phase (at $T<T_c^L$) \cite{MC}. In a sufficiently large flat trap
in which the condition $\tilde\epsilon^L>\tilde\epsilon^{\infty}/1.2$ is hold
\cite{eL>eoo/1.2} VU superfluid density $n_l$ thoughout the superfluid phase
obeys to the inequality $n_t>n_l>n^*$, where
$n^*\equiv n_l(T_c^L)=(0.93\div0.98)n_t$ for GaAs CQWs studied experimentally
in Refs. \cite{Toldnew,na417047,b6400413,rl962202,rl970103,b7400409,liqu,B})
and $n^*\approx0.9n_t$ in Refs. \cite{ss134037,ringS}. So, $n_l$ is close to
the total exciton density $n_t$ (for pure CQWs) \cite{MC}. Thus, for
spatially indirect CQW excitons we can assume VU superfluid exciton density
$n_l$ to be approximately equal to total one $n_t$ throughout the superfluid
phase
\begin{equation}\label{nlnt}
n_l\approx n_t\;\;\;\;(T<T_c^L).
\end{equation}
This permits not only to simplify Eqs. (\ref{ns}), (\ref{a}) and (\ref{_TcL})
\begin{equation}\label{nsa}
n_s=\frac{n_t}{\epsilon(L)},\;\;\;\;
a=\frac{2\pi\hbar^2n_t}{mT}=\frac{T_0}T,
\end{equation}
\begin{equation}\label{TcL}
T_c^L=\frac{\pi\hbar^2n_t}{2m\tilde\epsilon^L}=\frac{T_0}{4\tilde\epsilon^L},
\end{equation}
but also to perform a complete calculation of the VR superfluid exciton
density. (In Eqs. (\ref{nsa}) and (\ref{TcL}) the quantity
$T_0=2\pi\hbar^2n_t/m$ is the degeneration temperature of the spin-polarized
excitons; if the excitons are not spin-polarized and have $g_{ex}$ spin
degrees, the real exciton degeneration temperature is
$2\pi\hbar^2n_t/g_{ex}m$, this quantity is in $g_{ex}$ times less than $T_0$;
$g_{ex}=4$ for GaAs.) At typical total exciton density ({\it in all} spin
degrees \cite{spin}) $n_t=2\cdot10^{10}$ cm${}^{-2}$ and exciton mass
$m=0.22m_0$ ($m_0$ is free electron mass) \cite{Toldnew} the indirect exciton
superfluid crossover temperature in 2D flat trap is estimated as $T_c^L=1.2$ K
for the trap size $L=5.6$ $\mu$m, $T_c^L=1.15$ K for $L=43$ $\mu$m,
$T_c^L=1.1$ K for $L=14$ cm and $T_c^L=1.077$ K for an infinite ($L=\infty$)
system (see Eqs. (\ref{eLcr}) and (\ref{TcL}), degeneration temperature being
equal to $T_{deg}=1.26$ K.

In Fig. 2 we show the temperature dependence of the (VR) superfluid exciton
density in 2D flat trap. The temperature unit is KT temperature of the 2D
infinite superfluid \cite{rl391201}
\begin{equation}\label{Tcoo}
T_c^{\infty}\equiv\left.T_c^L\right|_{L=\infty}=
\frac{\pi\hbar^2n_t}{2m\tilde\epsilon^{\infty}}=
\frac{T_0}{4\tilde\epsilon^{\infty}},
\end{equation}
(here, as in Eq. (\ref{TcL}), we assume $n_l(T_c^{\infty})=n_t$; see Eq.
(\ref{nlnt})). We see that the superfluid density universal jump
$\tilde n_s^{\infty}=2mT_c^{\infty}/\pi\hbar^2=n_t/\tilde\epsilon^{\infty}$ is
{\it smeared out} in a flat trap as compared to infinite systems. Besides
smearing, the confinement affects in {\it growth} of the superfluid crossover
temperature.

The dependence of superfluid crossover temperature $T_c^L$ on trap size $L$
are depicted in Fig. 1. We see the fit (\ref{eLcr}) is much more better
coincides with the exact KT-model numerical intergation of Eq. (\ref{dex/dx}),
than the Kosterlitz's renormalization group calculation
\cite{jc071046}
\begin{equation}\label{eLcrRG}
\tilde\epsilon^L_{RG}=\left(1-\frac{\pi^2b^2}{\ln^2(L/r_0)}\right)
\tilde\epsilon^{\infty}
\end{equation}
(although, in the strict sence, Eqs. (\ref{eLcr}) and (\ref{eLcrRG}) coincide
in the logarithmic approximation).

\section{Superfluidity of 2D excitons in a large-size harmonic trap: The
generalized local density approximation}\label{GLDA}
The local density approximation for trapped excitons described in Sect.
\ref{LDA} is valid only if the total exciton density varies sufficiently
slowly on both typical phonon and vortex scales of the trapped superfluid.
But the typical vortex scales of 2D homogeneous infinite superfluid have the
{\it divergence} near KT transition \cite{jc071046,rl400783}. So, close to KT
transition the vortex scales cannot be much smaller than the (finite) trap
size (TF radius) $\bar L$. Moreover, at sufficiently low temperatures the
density of (thermally created) vortex pairs is negligible throughout the trap
except the narrow region near the boundary of the system. Thus, for more
accurate description of the exciton superfluidity in 2D inhomogeneous traps
the usual LDA must be modified.

The generalization of KT rare vortex pair gas theory for {\it inhomogeneous}
systems is not a simple problem. Indeed, vortex charge $q$ being proportional
to square root $\sqrt{n_l({\bf R})}$ of VU superfluid exciton density
$n_l({\bf R})$ \cite{jc061181} {\it varies} as trap point ${\bf R}$ displaces.
So, vortex pairs are no longer neutral dipoles. This result in an
inapplicability, in the strict sense, of KT theory for a inhomogeneous case.
However, this theory describes correctly the superfluid on the small length
scales (much smaller than trap size $\bar L$), on which the vortex charge
varies slowly. On the scales of order of $\bar L$ the integration of Eq.
(\ref{dex/dx}) must be effectively cutted off. But as in a flat trap, in a
harmonic trap vortex properties depend logarithmically on the length scale.
So, at sufficiently large $\bar L$ ($\ln(\bar L/r_0)\gg1$) in the logarithmic
approximation the exact value of the cutoff in integrating Eq. (\ref{dex/dx})
is not impotant. This results in appearing a possibility to investigate
quantitatively CQW exciton superfluid in a large-size harmonic trap with a
logarithmically small error in the approximation which we will call {\it
generalized} local density approximation (GLDA) for KT theory of weakly
inhomogeneous systems. In GLDA, VR exciton superfluid properties in 2D
large-size harmonic trap near to trap point ${\bf R}$ are replaced by ones in
2D {\it flat trap} with total density $n_t$ equal to total density
$n_t({\bf R})$ in harmonic trap, flat trap size $L$ being of order of harmonic
trap size $\bar L$.

We obtain readily the exciton fluid in 2D {\it symmetrical} harmonic trap at
low temperatures to divide in GLDA, as in LDA, into the superfluid circle
\begin{equation}\label{SCL}
n_t(R)>\tilde n_t^L,\;\;\;\;n_s(R)\sim n_t(R)\;\;\;\;(R<R_s^L)
\end{equation}
and normal ring
\begin{equation}\label{NRL}
n_t(R)<\tilde n_t^L,\;\;\;\;n_s(R)\approx0\;\;\;\;(R_s^L<R<\bar L).
\end{equation}
Here $R_s^L$ is the superfluid circle radius in the harmonic trap of size
$\bar L$, and (see Eq. (\ref{TcL}))
\begin{equation}\label{ntLcr}
\tilde n_t^L=\frac{2m\tilde\epsilon^L}{\pi\hbar^2}T
\end{equation}
is (critical) total exciton density in 2D flat trap of size $L\sim\bar L$, at
which excitons undergo the superfluid crossover at temperature $T$.

But near the circumference $R\approx R_s$, where $n_t(R)\approx\tilde n_t^L$,
which separates the superfluid circle and normal ring, there is a narrow
region of the local KT {\it crossover}, on which the onset of the pair vortex
dissociation and free vortex formation is {\it gradual}. This is one of
distinctions between LDA and GLDA.

The critical total density (\ref{ntLcr}) grows with the exciton temperature.
So, superfluid circle radius $R_s^L$ drops, the superfluid circle shrinks, and
the normal ring becomes wider (see Eqs. (\ref{SCL})-(\ref{ntLcr})). At the
temperature (see Eq. (\ref{ntLcr}))
\begin{equation}\label{_TcLt}
\bar T_c^L=\frac{\pi\hbar^2n_t^m}{2m\tilde\epsilon^L}
\end{equation}
at which the critical total exciton density $\tilde n_t^L$ is equal to the
maximal one in whole the trap $n_t^m=n_t(0)$ the excitons undergoes the global
superfluid {\it crossover} where VR superfluid component vanishes (in contrast
to LDA) {\it gradually}. Above $\bar T_c^L$ there is no superfluid circle, so
the trapped excitons are in globally normal phase.

We describe quantitatively the superfluidity of 2D CQW excitons in a
large-size harmonic trap in GLDA for TF density profile \cite{LV,MC}
$$
n_t(R)=n_l(R)=n_t^m(1-R^2/\bar L^2)\theta(\bar L-R),
$$
where $\theta(x)=0$ if $x<0$ and $\theta(x)=1$ if $x>0$. In this case, we
obtain in GLDA the following equations for superfluid circle radius $R_s^L$,
global superfluid crossover temperature $\bar T_c^L$, TF radius $\bar L$,
exciton chemical potential $\mu$, maximal total exciton density $n_t^m$ (in
trap center), and exciton numbers in the superfluid circle $N_{SC}$ and
normal ring $N_{NR}$:
$$
R_s^L=\bar L\sqrt{1-\tilde n_t^L/n_t^m}=\bar L\sqrt{1-T/\bar T_c^L},
$$
$$
\bar T_c^L=\frac{\hbar\omega_o\sqrt{N_t}}{2\tilde\epsilon^L\sqrt{2\zeta'}},
$$
$$
\bar L=x_o\sqrt[\mbox{\normalsize 4}]{8\zeta'N_t},
$$
$$
\mu=\hbar\omega_o\sqrt{2\zeta'N_t},
$$
$$
n_t^m=\frac{\sqrt{N_t/2\zeta'}}{\pi x_o^2},
$$
$$
N_{SC}=N_t(1-(T/\bar T_c^L)^2)\theta(\bar T_c^L-T),
$$
$$
N_{NR}=N_t((T/\bar T_c^L)^2\theta(\bar T_c^L-T)+\theta(T-\bar T_c^L)),
$$
where $\omega_o$ is the oscillator (trap) frequency,
$x_o=\sqrt{\hbar/m\omega_o}$ is the oscillator length, $N_t$ is the total
exciton number in the trap (in all spin degrees \cite{spin}), and
$\zeta'=\mu m/2\pi\hbar^2n_t^m=const$ is dimensionless exciton chemical
potential, which we assume to be independent of $T$ and $n_t^m$. Local and
global superfluid crossover halfwidths determining the correctness of the
logarithmic approximation (and hence, GLDA), as in a flat trap, have the third
order on the logaritmically small quantity, i.~e.,
$$
\Delta R_s^L\sim\frac{\bar L^2T}{2R_s^L\bar T_c^L}
\frac{\pi^2b^2}{(\ln(\bar L/r_0)+\Delta)^3},
$$
$$
\Delta\bar T_c^L\sim\bar T_c^L\frac{\pi^2b^2}{(\ln(\bar L/r_0)+\Delta)^3}.
$$

\begin{figure}[t]
\includegraphics[width=8.65cm,height=8.2cm]{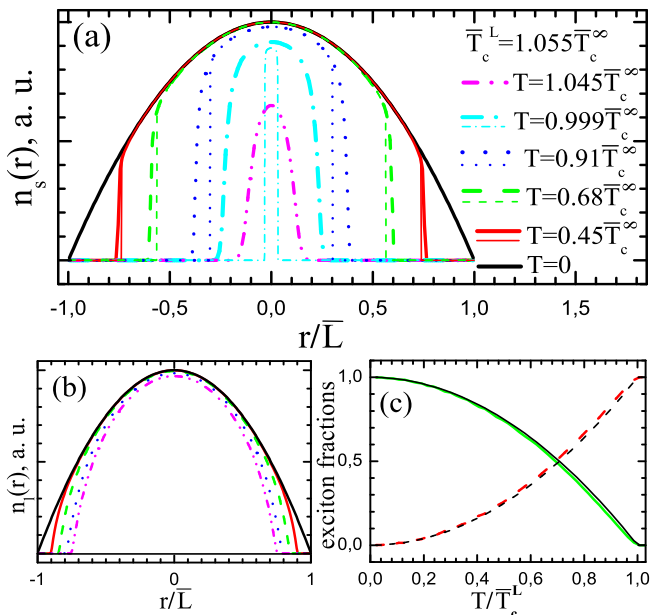}
\vskip -4mm
\caption{\small (a) The vortex-renormalized superfluid profiles of 2D CQW
excitons in the harmonic trap of large radius $\bar L$ ($\ln(\bar L/r_0)=8$)
at different temperatures $T$ in GLDA (thick lines) and in LDA (thin lines).
(b) The vortex-unrenormalized superfluid profiles in CQW structure (studied
experimentally in Ref. \cite{revT,Toldnew}) for $n_t^m=2.5\cdot10^{10}$
cm${}^{-2}$ at the same temperatures for corresponding lines as in (a). (c)
Solid lines shows the temperature dependence of superfluid (thick) and
superfluid circle (thin) fractions. Dash lines denotes normal (thick) and
normal ring (thin) fractions.}
\end{figure}

In Fig. 3 we show (VR) superfluid profiles for 2D harmonically trapped CQW
excitons in GLDA and in LDA (see Fig. 3(a)) at different temperatures below
$\bar T_c^L$ (the temperature unit is the global superfluid transition
temperature in LDA $\bar T_c^{\infty}\equiv\left.\bar T_c^L\right|_{L=\infty}=
(\tilde\epsilon^L/\tilde\epsilon^{\infty})\bar T_c^L<\bar T_c^L$; see Eqs.
(\ref{eLcr}), (\ref{TcL}), (\ref{Tcoo}) and (\ref{_TcLt})). VU superfluid
profiles are presented in Fig. 3(b) for CQW structure (see, e.~g.,
\cite{revT,Toldnew}) with exciton mass $m=0.22m_0$, dielectric constant
$\varepsilon=12.5$, and effective electron-hole layer distance $d=13.6$ nm for
trap center total exciton density being equal to $n_t^m=2.5\cdot10^{10}$
cm${}^{-2}$ \cite{MC}. One see the superfluid density on the superfluid
circle boundary region to drop sharply far from the global superfluid
transition ($T\not\approx\bar T_c^L$). Besides, there is little normal
component in the superfluid circle at $T\not\approx\bar T_c^L$ (for a {\it
pure} CQW structure). The latter fact is visually depicted in Fig. 3(c),
where the curve of the superfluid (normal) exciton number in GLDA quite merge
with that of exciton number in the superfluid circle (normal ring).

Because the global superfluid crossover temperature $\bar T_c^L$ for the
harmonic trap of radius $\bar L$ coincides with the temperature $T_c^L$ for
the flat trap of size $L\sim\bar L$ (see above), the dependence of the former
on $\bar L$ coincides with that presented in Fig. 1 with the change
$T_c^L\to\bar T_c^L$, $T_c^{\infty}\to\bar T_c^{\infty}$, and $L\to\bar L$.

\section{Conclusion}
We have calculated the temperature dependence of superfluid density and the
superfluid crossover temperature for 2D spatially indirect CQW excitons in a
large-size flat trap as well as superfluid profiles and global superfluid
crossover temperature for the excitons in a large-size harmonic trap. We
reveal that harmonically trapped exciton system at low temperatures is divided
into the superfluid circle (with superfluid component) and normal ring
(without it). At their boundary there is the narrow crossover region at which
the onset of vortex pair dissociation and free vortex formation takes place.
The superfluid density goes rather {\it sharply} to zero in this region. As
exciton temperature grows superfluid circle shrinkes to zero. This temperature
corresponds to the global superfluid transition (crossover) in the harmonic
trap. At this temperature the Bose condensate also disappears (details will be
published elsewhere).

\section*{Acknowledgements}
Authors are thankful to P. Littlewood, L. V. Keldysh, R. Zimmermann, and D.
Snoke for useful discussions of the results. The work was supported by the
Russian Foundation of Basic Research and Swedish Foundation for Strategic
Research (SSR).


\begin{thebibliography}{99}
\bibitem{LY}
Yu.E. Lozovik, V.I. Yudson, JETP Lett. 22 (1975) 274; JETP 44 (1976) 389;
Solid State Commun. 19 (1976) 391; {\it ibid.} 21 (1977) 211; Yu.E. Lozovik,
V.N. Nishanov, Phys. Solid State 18 (1976) 1905; Yu.E. Lozovik, A.M. Ruvinsky,
JETP 85 (1997) 979; Yu.E. Lozovik, Report on Conf. Diel. Electr., FAN, 1973.

\bibitem{SQW}
A. Filinov, P. Ludwig, Yu. E. Lozovik, M. Bonitz, H. Stolz, J. Phys: Conf.
Series 35 (2006) 197; P. Ludwig, A. Filinov, M. Bonitz, H. Stolz, Phys. Status
Solidi B 243 (2006) 2363.

\bibitem{SFB}
O.L. Berman, Yu.E. Lozovik, D.W. Snoke, R.D. Coalson, Phys. Rev. B 70 (2004)
235310; {\it ibid.} 73 (2006) 235352; Solid State Commun. 134 (2005) 47;
Physica E 34 (2006) 268; J. Phys.: Condens. Matter 19 (2007) 386219.

\bibitem{LB}
Yu.E. Lozovik, O.L. Berman, JETP Lett. 64 (1996) 573; JETP 84 (1997) 1027;
Phys. Scripta 55 (1997) 491; Yu.E. Lozovik, O.L. Berman, V.G. Tsvetus, {\it
ibid.} 66 (1997) 355; Phys. Rev. B 59 (1999) 5627; Yu.E. Lozovik, O.L. Berman,
M. Willander, J. Phys.: Condens. Matter 14 (2002) 12457; Yu. E. Lozovik, S. A.
Verzakov, M. Willander, Phys. Lett. A 260 (1999) 400; Yu.E. Lozovik, M.
Willander, Appl. Phys. A 71 (2000) 379.

\bibitem{BEC&SF}
G. Vignale, A.H. MacDonald, Phys. Rev. Lett. 76 (1996) 2786; X. Zhu, P.B.
Littlewood, M.S. Hibertsen, T.M. Rice, {\it ibid.} 74 (1995) 1633; Y. Naveh,
B. Laikhtman, {\it ibid.} 77 (1996) 900; J. Fern\'andez-Rossier, C. Tejedor,
{\it ibid.} 78 (1997) 4809; A.V. Balatsky, Y.N. Joglekar, P.B. Littlewood,
{\it ibid.} 93 (2004) 266801; Y.N. Joglekar, A.V. Balatsky, M.P. Lilly, Phys.
Rev. B 72 (2005) 205313; S. Conti, G. Vignale, A.H. MacDonald, {\it ibid.} 57
(1998) 6846; J.F. Jan, Y.C. Lee, {\it ibid.} 58 (1998) 1714; A.L. Ivanov, P.B.
Littlewood, H. Haug, {\it ibid.} 59 (1999) 5032; S.-R.E. Yang, J. Yeo, S. Han,
{\it ibid.} 71 (2005) 245307; A.L. Ivanov, Europhys. Lett. 59 (2002) 586.

\bibitem{LV}
Yu.E. Lozovik, S.Yu. Volkov, M. Willander, JETP Lett. 79 (2004) 473; G.E.
Astrakharchik et al., arXiv:0707.4630.

\bibitem{cr}
Yu.E. Lozovik, O.L. Berman, Phys. Scripta 58 (1998) 86; Phys. Solid State 40
(1998) 1228; D.V. Kulakovsky, Yu.E. Lozovik, A.V. Chaplik, JETP 99 (2004) 850.

\bibitem{MCcr}
G.E. Astrakharchik et al., Phys. Rev. Lett. 98 (2007) 060405; H.P. B\"uchler
et al., {\it ibid.} 98 (2007) 060404; S. De Palo, F. Rapisarda, G. Senatore,
{\it ibid.} 88 (2002) 206401; Yu.E. Lozovik, E.A. Rakoch, Phys. Lett. A 235
(1997) 55; A.I. Belousov, Yu.E. Lozovik, JETP Lett. 68 (1998) 858; Eur. Phys.
J. D 8 (2000) 251; A. Filinov, M. Bonitz et al., Phys. Status Solidi C 3
(2006) 2457; C. Mora, O. Parcollet, X. Waintal, Phys. Rev. B 76 (2007) 064511.

\bibitem{YKSP}
Yu.E. Lozovik, V.I. Yudson, Solid State Commun. 22 (1977) 117; A.V. Klyuchnik,
Yu.E. Lozovik, JETP 49 (1979) 335; J. Phys. C 11 (1978) L483; S.I. Shevchenko,
Phys. Rev. Lett. 72 (1994) 3242; Yu.E. Lozovik, A.V. Poushnov, Phys. Lett. A
228 (1997) 399.

\bibitem{opt}
B. Laikhtman, Europhys. Lett. 43 (1998) 53; P. Stenius, Phys. Rev. B 60
(1999) 14072; Yu.E. Lozovik, I.V. Ovchinnikov, {\it ibid.} 66 (2002) 075124;
JETP Lett. 74 (2001) 288; Solid State Commun. 118 (2001) 251; Yu.E. Lozovik,
I.L. Kurbakov, I.V. Ovchinnikov, {\it ibid.} 126 (2003) 269; R. Zimmermann
{\it ibid.} 134 (2005) 43; A. Olaya-Castro, F.J. Rodr\'iguez, L. Quiroga, C.
Tejedor, Phys. Rev. Lett. 87 (2001) 246403; J. Keeling, L.S. Levitov, P.B.
Littlewood, {\it ibid.} 92 (2004) 176402; L.S. Levitov, B.D. Simons, L.V.
Butov, {\it ibid.} 94 (2005) 176404.

\bibitem{revM}
S.A. Moskalenko, D.W. Snoke, Bose-Einstein condensation of excitons and
biexcitons and coherent nonlinear optics with excitons, Cambridge University
Press, Cambridge, 2000.

\bibitem{revT}
V.B. Timofeev, Phys.-Uspekhi 48 (2005) 295; V.B. Timofeev, A.V. Gorbunov, A.V.
Larionov, J. Phys.: Condens. Matter 19 (2007) 295209.

\bibitem{revB}
L.V. Butov, J. Phys.: Condens. Matter 16 (2004) R1577; {\it ibid.} 19 (2007)
295202.

\bibitem{revS}
D. Snoke, Science 298 (2002) 1368; R. Rapaport, G. Chen, J. Phys.: Condens.
Matter 19 (2007) 295207; Z. V\"or\"os et al., {\it ibid.} 19 (2007) 295216.

\bibitem{KK}
L.V. Keldysh, Y.V. Kopaev, Phys. Solid State 6 (1965) 2219; A.N. Kozlov, L.A.
Maksimov, JETP 21 (1965) 790; L.V. Keldysh, A.N. Kozlov, {\it ibid.} 27
(1968) 521.

\bibitem{r1580383}
P.C. Hohenberg, Phys. Rev. 158 (1967) 383; N.D. Mermin, H. Wagner, Phys. Rev.
Lett. 17 (1966) 1133.

\bibitem{r1550080}
J.W. Kane, L.P. Kadanoff, Phys. Rev. 155 (1967) 80.

\bibitem{Berezinskii}
V.L. Berezinskii, JETP 32 (1970) 493; {\it ibid.} 34 (1971) 610.

\bibitem{jc061181}
J.M. Kosterlitz, D.J. Thouless, J. Phys. C 6 (1973) 1181.

\bibitem{jc071046}
J.M. Kosterlitz, J. Phys. C 7 (1974) 1046.

\bibitem{2DBEC}
V. Bagnato, D. Kleppner, Phys. Rev. A 44 (1991) 7439; W. Ketterle, N.J. van
Druten, {\it ibid.} 54 (1996) 656.

\bibitem{Toldnew}
A.V. Gorbunov, V.E. Bisti, V.B. Timofeev, JETP 101 (2005) 693; A.V. Larionov,
V.B. Timofeev el al., {\it ibid.} 90 (2000) 1093; JETP Lett. 71 (2000)
117; {\it ibid.} 75 (2002) 200; {\it ibid.} 75 (2002) 570; A.V. Larionov, V.B.
Timofeev, {\it ibid.} 73 (2001) 301; A.A. Dremin, V.B. Timofeev, A.V. Larionov
et al., {\it ibid.} 76 (2002) 450; A.V. Gorbunov, V.B. Timofeev, {\it ibid.}
83 (2006) 146; {\it ibid.} 84 (2006) 329; A.V. Gorbunov, A.V. Larionov, V.B.
Timofeev, {\it ibid.} 86 (2007) 46; V.B. Timofeev, A.V. Gorbunov, J. Appl.
Phys. 101 (2007) 081708; A.A. Dremin, A.V. Larionov, V.B. Timofeev, Phys.
Solid State 46 (2004) 170.

\bibitem{na417047}
L.V. Butov et al., Nature (London) 417 (2002) 47.

\bibitem{b6400413}
V.V. Krivolapchuk, E.S. Moskalenko, A.L. Zhmodikov, Phys. Rev. B 64 (2001)
045313.

\bibitem{rl962202}
A.T. Hammack et al., Phys. Rev. Lett. 96 (2006) 227402.

\bibitem{rl970103}
Z. V\"or\"os et al., Phys. Rev. Lett. 97 (2006) 016803.

\bibitem{ss134037}
D.W. Snoke et al., Solid State Commun. 134 (2005) 37.

\bibitem{a0990604}
A.T. Hammack et al., J. Appl. Phys. 99 (2006) 066104.

\bibitem{mj360940}
M. Willander, O. Nur, Yu.E. Lozovik et al., Microelectron. J. 36 (2005) 940.

\bibitem{b7400409}
G. Chen et al., Phys. Rev. B 74 (2006) 045309.

\bibitem{ringS}
D. Snoke et al., Nature (London) 418 (2002) 754; Solid State Commun. 127
(2003) 187; R. Rapaport, G. Chen, D. Snoke et al., Phys. Rev. Lett. 92 (2004)
117405.

\bibitem{liqu}
V.V. Krivolapchuk, E.S. Moskalenko, A.L. Zhmodikov et al., Solid State Commun.
111 (1999) 49; Phys. Solid State 41 (1999) 291.

\bibitem{B}
L.V. Butov et al., Nature (London) 418 (2002) 751; Phys. Rev. B 58 (1998)
1980; {\it ibid.} 59 (1999) 1625; Phys. Rev. Lett. 73 (1994) 304; {\it ibid.}
86 (2001) 5608; {\it ibid.} 92 (2004) 117404; S. Yang et al, {\it ibid.} 97
(2006) 187402; C.W. Lai et al., Science 303 (2004) 503.

\bibitem{LDAcorr}
LDA for KT theory was also discussed in talk of Yu.E.L. on the Cambridge
Workshop on exciton BEC, 2004.

\bibitem{rl391201}
D.R. Nelson, J.M. Kosterlitz, Phys. Rev. Lett. 39 (1977) 1201.

\bibitem{b3203088}
P. Minnhagen, Phys. Rev. B 32 (1985) 3088.

\bibitem{VR}
P. Minnhagen and G.G. Warren, Phys. Rev. B 24 (1981) 2526; P. Minnhagen, Rev.
Mod. Phys. 59 (1987) 1001.

\bibitem{MC}
Yu.E. Lozovik et al., JETP, in press (2008); Solid. State Commun. 144 (2007)
399.

\bibitem{TT}
Yu.E. Lozovik, I.L. Kurbakov, M. Willander (to be publ.).

\bibitem{eL>eoo/1.2}
The condition $\tilde\epsilon^L>\tilde\epsilon^{\infty}/1.2$ corresponds to
sizes $L$, obeyed the inequality $\ln(L/r_0)>3.2$ (see Fig. 1 and Eq.
(\ref{eLcr})), i.~e., it takes place for the flat traps with more than
$(L/r_0)^2\sim600$ excitons.

\bibitem{spin}
Throughout this Letter all densities and particle numbers are given in all
spin degrees. We make this denotion because it is the density in all spin
components that determines VR superfluid effects. This is associated with that
a vortex rotates all $g_{ex}$ spin components.

\bibitem{rl400783}
V. Ambegaokar, B.I. Halperin, D.R. Nelson, E.D. Siggia, Phys. Rev. Lett.
40 (1978) 783.
\end{thebibliography}
\end{document}